\documentclass{webofc}
\usepackage[varg]{txfonts}   % Web of Conferences font
\usepackage{lineno}
\usepackage{natbib}
\begin{document}
\title{Evolution of the energy efficiency of LHCb's real-time processing}

\author{\firstname{Roel} \lastname{Aaij}\inst{1} \and
        \firstname{Daniel Hugo} \lastname{C\'{a}mpora~P\'{e}rez} \inst{2} \and
        \firstname{Tommaso} \lastname{Colombo}\inst{3} \and
        \firstname{Conor} \lastname{Fitzpatrick}\inst{4} \and
        \firstname{Vladimir Vava} \lastname{Gligorov}\inst{5}\fnsep\thanks{\email{vgligoro@lpnhe.in2p3.fr}} \and
        \firstname{Arthur} \lastname{Hennequin}\inst{3,6} \and
        \firstname{Niko} \lastname{Neufeld}\inst{3} \and
        \firstname{Niklas} \lastname{Nolte}\inst{7} \and
        \firstname{Rainer} \lastname{Schwemmer}\inst{3}\fnsep\thanks{\email{Rainer.Schwemmer@cern.ch}} \and
        \firstname{Dorothea} \lastname{Vom Bruch}\inst{8}
        % etc.
}
\institute{
Nikhef National Institute for Subatomic Physics, Amsterdam, Netherlands
\and
Universiteit Maastricht, Maastricht, Netherlands
\and
European Organization for Nuclear Research (CERN), Switzerland
\and
Department of Physics and Astronomy, University of Manchester, Manchester, United Kingdom
\and
           LPNHE, Sorbonne Universit{\'e}, CNRS/IN2P3, France
\and
           LIP6, Sorbonne Universit{\'e}, France
\and
Massachusetts Institute of Technology, Cambridge, MA, United States
\and
Aix Marseille Univ, CNRS/IN2P3, CPPM, Marseille, France
          }

\abstract{%
  The upgraded LHCb detector, due to start datataking in 2022, will have to process an average data rate of
  4~TB/s in real time. Because LHCb's physics objectives require that the full detector information for
  every LHC bunch crossing is read out and made available for real-time processing, this bandwidth challenge 
  is equivalent to that of the ATLAS and CMS HL-LHC software read-out, but deliverable five years earlier.
  Over the past six years, the LHCb collaboration has undertaken a bottom-up rewrite of its software
  infrastructure, pattern recognition, and selection algorithms to make them better able to efficiently
  exploit modern highly parallel computing architectures. We review the impact of this reoptimization
  on the energy efficiency of the real-time processing software and hardware which will be used for the
  upgrade of the LHCb detector. We also review the impact of the decision to adopt a hybrid computing
  architecture consisting of GPUs and CPUs for the real-time part of LHCb's future data processing. We discuss
  the implications of these results on how LHCb's real-time power requirements may evolve in the future,
  particularly in the context of a planned second upgrade of the detector.
}
\maketitle
\section{Introduction}
\label{sec:intro}
\begin{figure}[t]
  \centering
  \includegraphics[width=0.95\textwidth]{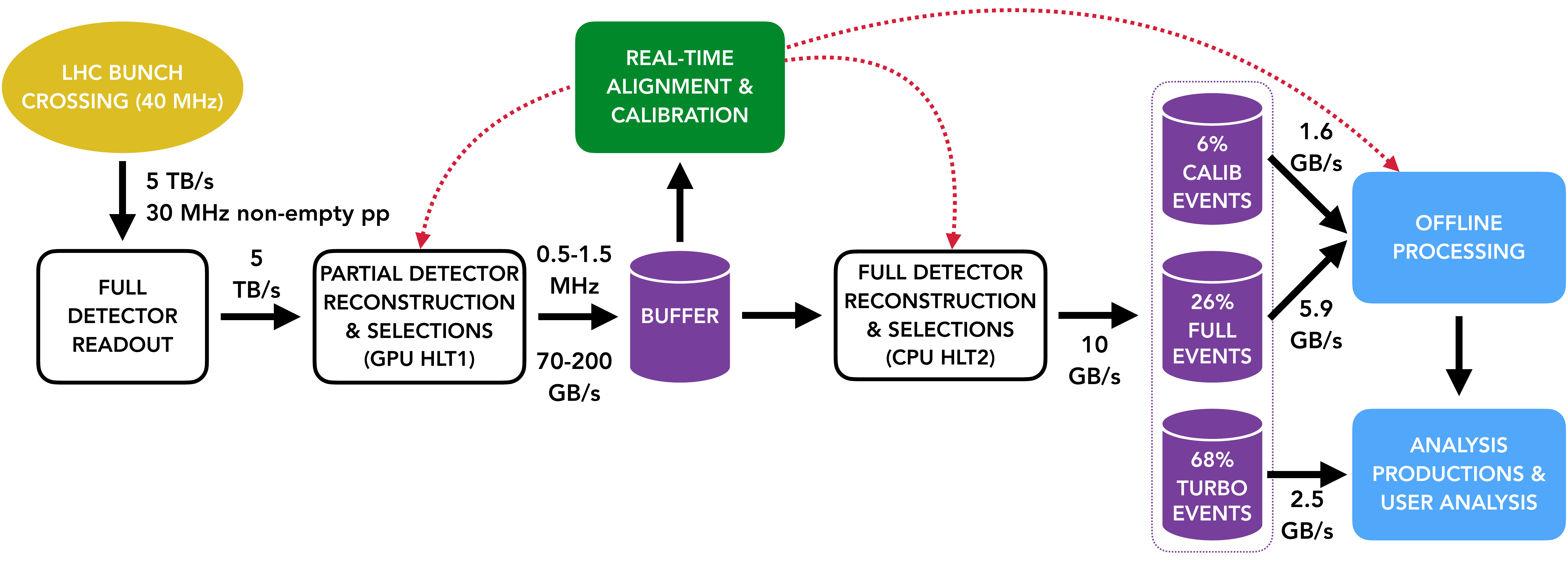}
  \caption{Dataflow in the upgraded LHCb detector, reproduced from~\cite{LHCb-FIGURE-2020-016}.}
  \label{fig:lhcbupdataflow}
\end{figure}

The upgrade of the LHCb detector, due to begin datataking in Run~3, will use a
triggerless readout to send the raw data for the full 30 MHz of LHC bunch crossings to a custom data
processing centre. Its dataflow is illustrated in Figure~\ref{fig:lhcbupdataflow}. With a peak data rate of
5~TB/s and an expected average rate of 4~TB/s, LHCb's upgrade will have to handle similar data rates as
the ATLAS and CMS HL-LHC software triggers, but around five years earlier, making this one of the most complex
data processing challenges in HEP today.

This data processing will use off-the-shelf components to first build the raw bunch
crossing fragments from each subdetector into a single ``event'' (Event Building), and subsequently to group
the detector hits into higher level physics objects (High Level Trigger or HLT) and decide whether these are of
sufficient interest to warrant further analysis of the event. Its requirements and design
were described in~\cite{LHCb-TDR-016}. The first HLT stage
(HLT1) performs a partial event reconstruction based on a relatively small subset of inclusive selection
algorithms, each of which selects a general topology of interest to a range of LHCb's physics analyses. The
events which receive a positive decision at HLT1 are then processed by HLT2, which performs a full
reconstruction and uses hundreds of selection algorithms to fully build and isolate signal candidates of
interest to specific LHCb analyses. The events which receive a positive decision at HLT2 are recorded to permanent storage and sent to the GRID for further processing and distribution to the analysts.

LHCb's real-time processing software is historically based on the GAUDI~\cite{BARRAND200145} framework, 
with a number of additional features related to data integrity and the bookkeeping of provenance information
required by the real-time use case. At the start of Run~1 the HLT ran a significantly simplified
version of the offline pattern recognition and event selection algorithms, even in HLT2.~\cite{LHCb-DP-2012-004} 
Over the course of Run~1 the available processing power was increased and the processing algorithms gradually made faster, so
that by Run~2 LHCb was able to deploy its full offline reconstruction in HLT2.~\cite{LHCb-DP-2019-001} 
This led to the so-called ``Real Time Analysis'' (RTA) processing model
which remains the baseline for the LHCb upgrade.~\cite{LHCb-DP-2016-001,LHCb-TDR-018}  

In the RTA model, events selected by HLT1 are buffered for long enough to perform an
offline-quality detector alignment and calibration. This in turn allows the full detector reconstruction
deployed in HLT2 to build complete signal candidates of interest, and then record only information strictly
necessary for the final physics analysis of those candidates, throwing away the raw detector information and
most of the intermediate pattern recognition objects. The RTA model was further refined during Run~2 
datataking to allow an analysis-specific subset of raw detector data or 
associated pattern recognition objects to be appended to each RTA signal candidate.~\cite{LHCb-DP-2019-002} This transformed 
the binary choice between the traditional offline and RTA analysis models into a spectrum, which each 
analyst could tune to balance the data rate per selected event in the way which best suited their analysis.

LHCb decided on the basic parameters of its upgrade real-time processing architecture, including cost, in 2014. At
the time CPUs were expected to rapidy evolve towards having dozens if not hundreds of cores. 
This guided LHCb's priorities: the software had to become thread-safe, and
the transient data classes had to become much lighter to fit better into the processor cache and optimally support
vectorized computations. In parallel LHCb hedged its bets by implementing various reconstruction
algorithms on GPUs and other highly parallel architectures. These efforts converged in 2020, with LHCb
choosing a GPU architecture for HLT1 while remaining with the traditional 
CPU architecture for HLT2.~\cite{Aaij:2019zbu,LHCb-TDR-021}

The fact that this hybrid architecture could be delivered within the available cost envelope is a
combination of the price evolution of the GPU and CPU architectures, the ability to exploit CPU processing
resources left over from Run~2, and performance gains made by LHCb's software developers over the past
years. Although this performance evolution has been documented in public notes, we summarise it here
in a coherent manner, making clear the importance of each component. We document for the first time
the impact of the new GPU and CPU architectures, and the performance gains of LHCb's software, on the power
consumption of the system. Finally, we discuss the implications of these developments on the future
evolution of LHCb's real-time processing, particularly in the context of a planned
second upgrade of the LHCb detector.~\cite{LHCb-PII-EoI,Aaij:2636441} 

\section{Evolution of LHCb's HLT performance}
\label{sec:softperf}

We will use HLT1 in order to illustrate how the performance of LHCb's 
software developed over the past years. As this part of the codebase has to handle the biggest processing throughput,
of around 30~MHz and 4~TB per second, it also places the most stringent requirements on the software framework and
algorithms, and exposes suboptimal implementations that could remain hidden in later processing steps where more
time is available. It is worth noting here that because of LHCb's budget constraints, HLT1 had to be implemented
on O(10$^4$) physical cores, which at 30~MHz requires a processing rate of 3~kHz from each core.
This is a good demonstration of how LHCb's asynchronous DAQ architecture,
and the use of deep memory buffers in the servers which receive the event data, allows to hide the latency inherent with
the use of CPU (and later GPU) architectures while operating in a per-bunch-crossing time budget typically 
associated with FPGAs.

\begin{figure}[t]
  \centering
  \includegraphics[width=0.95\textwidth]{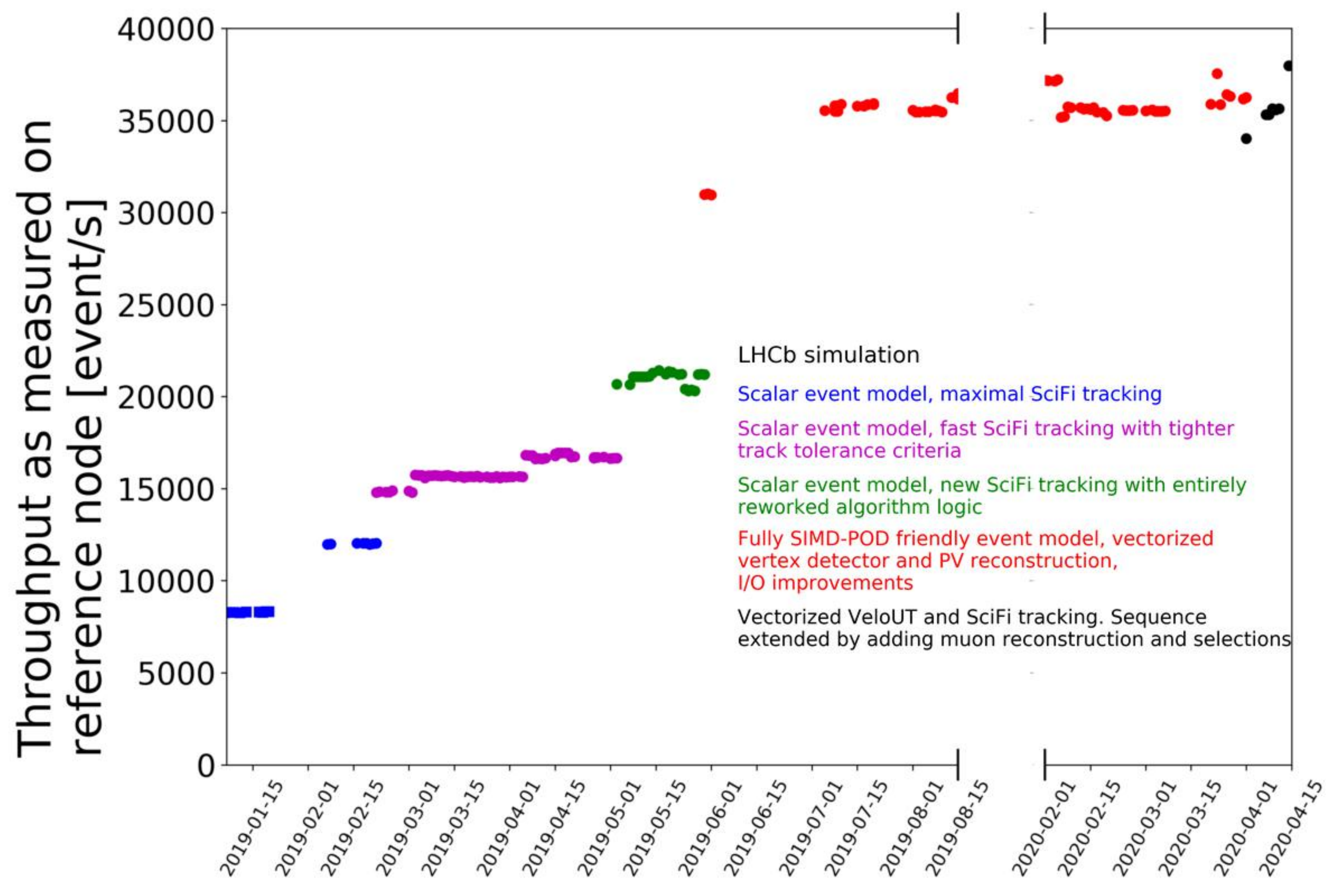}
  \caption{The evolution of HLT1 throughput as measured on a reference dual-socket E5-2630-v4 Intel Xeon node, reproduced from~\cite{LHCB-FIGURE-2020-007}. Expected pileup per bunch: 6.}
  \label{fig:lhcbcputhroughputevolution}
\end{figure}

The main goal of the HLT1 trigger is to find charged particles (tracks) in LHCb's vertex detector and the tracking system
downstream of the LHCb magnet, combine them into charged particle trajectories, and identify particularly interesting
one- and two-track combinations in order to decide which events to keep for further processing. Muon identification is
used to allow looser kinematic and geometric selection criteria to be applied in muonic final states.
The evolution of the HLT1 CPU throughput is summarized in Figure~\ref{fig:lhcbcputhroughputevolution}. Although
the plotting begins at the start of 2019, the gains made during 2019 were of course in large part due to preparatory
work done between 2016 and 2019 to which it is difficult to do justice on a plot, since many of the components had
to come together before really significant improvements could be seen. Nevertheless, the plot and its legend summarise
the essential steps in the process. The achieved gains were a combination of three complementary development directions:

\begin{enumerate}
    \item The ``physics'' reoptimization of algorithms, for example tightening search windows and selection criteria in 
    the charged particle reconstruction, and using a simplified straight-line track fit instead of a Kalman fit. These
    changes gained about a factor two in throughput performance without significant losses in the physics performance, but
    would not on their own have been enough to make the system viable for Run~3.
    \item Reoptimization of the transient data structures inspired by the PODIO~\cite{Gaede:2017ctv} paradigm. This work
    involved the removal of unnecessary pointer chasing, a switch to SIMD-friendly structure-of-arrays format wherever
    possible, and the removal of unnecessary data members to minimize memory use and fit better into the CPU cache. 
    \item A complete rewrite of the key tracking algorithms in order to vectorize the majority of time-consuming
    computational steps, taking advantage of the lighter and SIMD-friendly data structures designed in the previous step.
    The combination of vectorized algorithms and lighter data structures gained over a factor two in performance on top
    of what was achieved through physics optimizations alone.
\end{enumerate}

In addition a dedicated optimization of the I/O configuration was required because of the extremely high per-core
throughputs which had to be sustained. Thanks to this optimization, throughputs 
of over 5~kHz or 500~MB per second per physical CPU core could be achieved without saturating the I/O as documented in~\cite{Hennequin:2019itm}. In particular
this work enabled the efficient use of AMD's new EPYC generation of CPUs, which were released in 2019 and offered
up to 64 physical cores per socket as compared to the 10 cores of the reference E5-2630-v4 Intel Xeon node which formed
the backbone of LHCb's Run~2 real-time processing infrastructure. The performance of LHCb's CPU HLT1 on the 7502~AMD~EPYC
with 32 physical cores per socket was also documented in~\cite{LHCB-FIGURE-2020-007}, reaching 171~kHz per 
dual-socketed server, a factor $4.5$ times faster than the reference dual-socket E5-2630-v4~Intel~Xeon node. As the TDP
of the 7502~EPYC is 180~Watts, while that of the E5-2630-v4~Xeon is 85~Watts, it seems apparent that the new generation
of servers would also enable a significant improvement in energy efficiency, as we will document quantitatively in the next
section.

In parallel to this reoptimization of its CPU software, LHCb had been pursuing R\&D activities into GPUs and other
highly parallel computing architectures since the early 2010s. In early 2018, since there was not yet a CPU HLT1 which
could fit into the available computing resources, it was decided to attempt to implement the full HLT1 on GPUs as
a hedge against the possibility that the CPU implementation would not make it in time. This project, codenamed
``Allen'', successfully concluded in 2020 and demonstrated sufficient benefits over the CPU implementation that LHCb
finally decided to switch its baseline to a GPU HLT1. This decision was taken despite the fact that by 2020 the optimized
CPU HLT1 discussed earlier could fit into the available computing resources, and the reasons for it have been documented
in more detail elsewhere.~\cite{comparison} In terms of software design both the CPU HLT1 and GPU HLT1 followed analogous design principles for efficient, parallel and architecture-aware programming. The latest documented performance of the GPU HLT1 can be found 
in~\cite{LHCB-FIGURE-2020-014}; for the purposes of this paper the relevant number is a throughput of 148~kHz on a reference
Nvidia GeForce~2080~Ti GPU with a TDP of 250~Watts. Here again we see that LHCb's asynchronous processing architecture allows a GPU to leverage its parallel processing capabilities so that a compact system of O(200) GPU cards, comparable to the number
of boards typically used in ``hardware'' first-level triggers, is capable of ingesting the full bunch crossing rate of 30 MHz.
The beauty of this system is that by eliminating latency from the equation we do not have to provide a decision within microseconds, but rather that we ``just'' have to manage a high average throughput.

\section{Impact on power consumption}
\label{sec:power}

The improvements in processing speed documented in the previous section also clearly imply an improvement in the
energy efficiency of the system, in other words the number of trigger decisions which can be taken per kWh of energy
consumed. While the improvement documented from the physics and software reoptimization of the CPU HLT1 can be easily
translated into a gain in energy efficiency, the improvement from using a more modern CPU or from using a GPU cannot be
deduced simply from the processor's TDP, as overheads related to the server, as well as software inefficiencies which not always allow to reach the full TDP, must also be taken into account. We have
therefore carried out a dedicated set of power consumption measurements for three reference architectures: a dual-socket
E5-2630-v4 Intel Xeon node\footnote{Quanta DA0S2SMBCE0}; a dual-socket 7502 AMD EPYC\footnote{Gigabyte G482-Z5} node; and the same dual-socket AMD 7502 EPYC node with between one
and three GeForce~RTX~2080~Ti\footnote{Gigabyte GV-N208TTURBO-11GC-rev-10} Nvidia GPUs. 

In the case of the two CPU architectures we take the total power consumption of the
server as the reference measurement. In the case of GPUs the calculation is a bit more difficult. The GPUs are hosted opportunistically in the same servers used for LHCb's real-time event building (Event Building Nodes), replacing two high-speed network cards
which would have otherwise been required to send data to a CPU HLT1 implemented in dedicated CPU servers. We therefore
take the difference between the power consumption of the server when the GPUs are idle and when they are under load,
and subtract the power consumption of the network cards which are no longer required to be present. The results are
presented in Table~\ref{tab:powerconsumption} under the network replacement heading. The table also contains the calculation for a dedicated GPU machine. I.e. instead of putting the GPUs into the Event Building Nodes, data would be sent to dedicated GPU servers which contain appropriate numbers of network cards and processing GPUs.

Compared to Run~2 LHCb has gained over an order of magnitude in energy 
efficiency, with the gains divided rather evenly between improvements in the physics logic of HLT1, improvements in
the underlying software architecture and the use of SIMD programming paradigms, and improvements from the use of newer
processing architectures such as AMD's EPYC CPUs or Nvidia GPUs. 

\begin{table}[t]
\caption{Energy per trigger decision, in Millijoule, of different HLT1 architectures. The gain and cumulative gain are given with respect to the previous row in the table. For the exclusive and pure GPU rows, the gain and cumulative gain in power consumption are given with respect to the 7502 EPYC architecture. For the GeForce~RTX~2080~Ti results, NR stands for ``network replacement''.}
\begin{center}
\label{tab:powerconsumption}       % Give a unique label
\begin{tabular}{lccc}
\hline
Architecture & Energy per trigger (mJ) & Gain & Total gain \\
\hline
E5-2630-v4 Xeon & & & \\
\phantom{ind}Before SW optimization & 39.9 & 1.0x & \\
\phantom{ind}w/Physics optimizations & 21.0 & 1.9x & 1.9x \\
\phantom{ind}w/SIMD optimizations & 8.4 & 2.5x & 4.8x\\
7502 EPYC & & & \\
\phantom{ind}w/SIMD optimizations & 3.2 & 2.6x & 12.5x \\
Event Building Node, NR & & & \\
\phantom{ind} 1 GPU & 3.1 & 1.03x & 12.9x \\
\phantom{ind} 2 GPUs & 2.4 & 1.29x & 16.6x \\
\phantom{ind} 3 GPUs & 2.1 & 1.15x & 19.0x \\
Dedicated GPU machine & & & \\
\phantom{ind} 4 x 2080 Ti + 2 Network Cards & 2.8 & 1.14x & 14.3x \\
\phantom{ind} 5 x 2080 Ti + 3 Network Cards & 2.5 & 1.12x & 16.0x \\
Pure GPU machine & & & \\
\phantom{ind} 8 x 2080 Ti + Onboard Network & 2.1 & 1.15x & 19.0x \\
\hline
\end{tabular}
\end{center}
\end{table}

The power measurements were performed using the server internal power monitoring of the baseboard management controller (BMC). The Xeon platform comes in the form factor of a quad node (4 servers sharing one power supply and cooling system). These high density servers are generally regarded as extremely power efficient. For a fair comparison we ran our software on all 4 machines and then normalized to a single node. The AMD EPYC platform for this test was a single unit chassis which is normally used as a GPU hosting machine for up to 8 cards. Since these chassis are less power efficient to cope with the expected GPU hardware inside, we tuned the cooling system (for the CPU-only runs) to a mode that provided only cooling for the CPUs. For the GPU and GPU + Event Building configurations the cooling system was tuned to for the requirements of the DAQ cards. While these cards consume less power than a GPU, their thermal management is not as sophisticated and requires much higher air flows than GPUs. Each test configuration on all platforms was run and averaged over at least 10 minutes to allow the values to stabilize. 

\begin{figure}[t]
  \centering
  \includegraphics[width=0.8\textwidth]{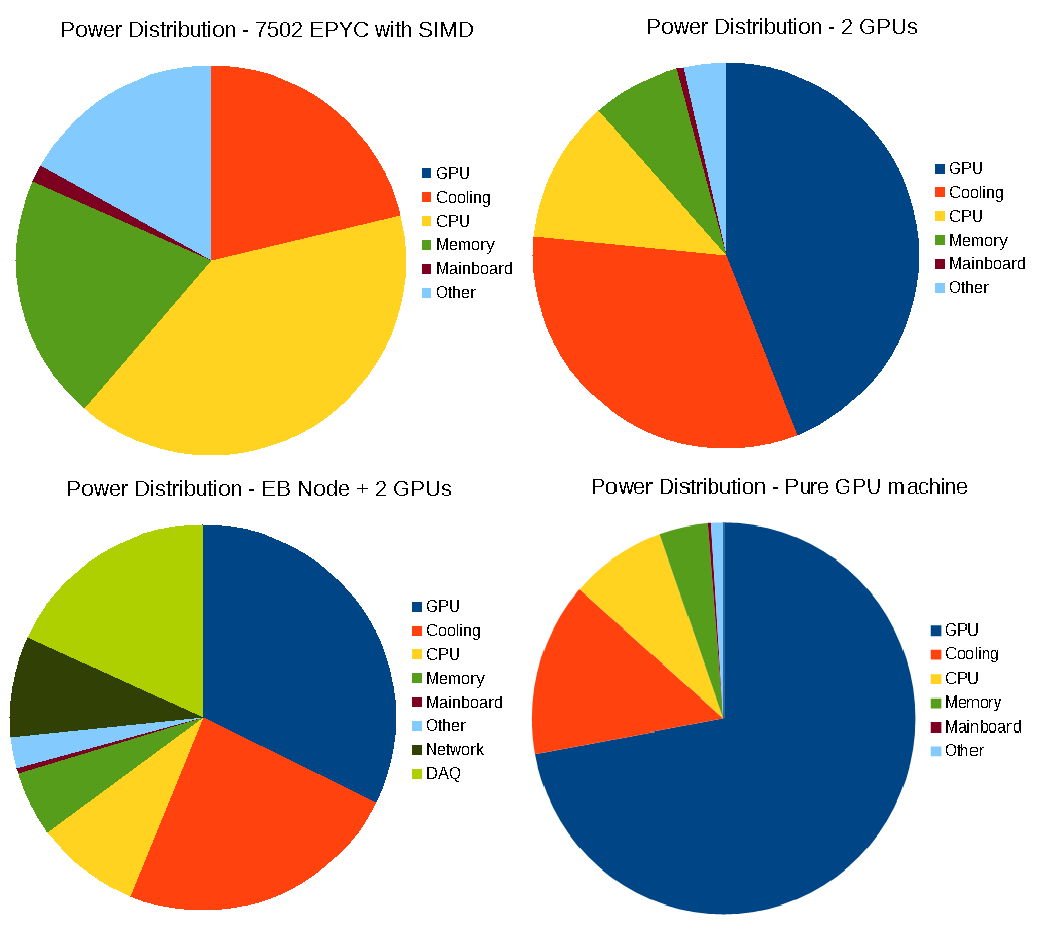}
  \caption{Power distribution of AMD EPYC platform according to the different extension card configurations described in the text.}
  \label{fig:powerbreakdown}
\end{figure}

Since the AMD platform provides excellent instrumentation for power measurements we are able to go into a bit more detail and Figure \ref{fig:powerbreakdown} provides a breakdown of the consumption of the various main components in the server. This break down tries to illustrates why there is such a wide range of power efficiency values for different GPU configurations.

\begin{itemize}
    \item \emph{7502 EPYC with SIMD}: The machine is only used to perform trigger decisions on the CPU, which was the candidate of the original CPU based trigger architecture. As expected, this configuration is dominated by CPU and Memory power consumption.
    \item \emph{2 GPUs}: The machine contains only 2 GPU cards which run the trigger. the CPU is mostly idle and is only orchestrating and monitoring the data flow to the GPU. This configuration contains no network cards and is dominated by cooling and the GPU. It is included here to illustrate the differences to the next configuration.
    \item \emph{Event Building Node + 2 GPUs}: The current candidate for the combined event builder + HLT1 trigger. While the GPUs consume about one third of the total power consumption, about a quarter of the power consumption goes into the DAQ task of the machine (event building network and DAQ FPGA cards). Note: The network depicted here is not the same network that the GPUs are replacing. This would be another slice, equivalent to the network part shown here. 
    \item \emph{Pure GPU machine (8 x 2080 Ti)}: This configuration is what the machine was originally designed to do. It is of very little use to us, since it does not have enough network bandwidth to feed the GPUs with data in our case.  It is used here illustrates the typical use case for a GPU focused processing node with lower throughput requirements.
\end{itemize}

Since cooling is such a large contributor in all configurations, it is important to take it into account for the efficiency calculations. In the case of the network replacement section of Table~\ref{tab:powerconsumption} we additionally distribute the power for cooling proportionally to the cards that are in the server. I.e. The \emph{EB Node + 2 GPUs} configuration contains 2 GPU cards, 2 event building network cards and 3 DAQ cards. Of the 390~W going into the chassis fans, only $\frac{2}{5}$ is  attributed to the GPU cards. While this might seem a bit unfair at first (the GPUs consume more power) the cooling requirements are solely driven by the DAQ cards. The heat spreaders on the DAQ cards are less efficient than the GPUs and the GPUs are actually overcooled in this configuration. Energy per compute efficiency could be at the \emph{8 GPUs} level if we could tune for the GPU requirements only.

\section{Conclusions and implications for the future}
\label{sec:implications}
The power consumption of real-time processing architectures has traditionally been a secondary consideration compared
to their physics performance and cost. However, as our experiments prepare to process ever bigger and more complex datasets in
decades which will be dominated by the fight against global warming, we must expect this situation to change. To
put the results of Table~\ref{tab:powerconsumption} in context, LHCb's detector and support electronics consume approximately 0.6 MW. LHCb's warm dipole magnet has a power consumption of 4.2~MW.
Had LHCb done no work to improve the efficiency of its software, the power consumption of an HLT1 implemented on Intel's E5-2630-v4 Xeon servers (which formed the backbone of LHCb's Run~2 real-time processing) would have exceeded 1.2~MW. 
If we add to this the power consumption of HLT2
as documented at the start of 2019 in~\cite{LHCB-FIGURE-2019-004}, which must process an input event rate of at least 1MHz, we would have reached an overall power consumption of over 4.8~MW, even without taking into account the overheads due to network
communications, the disk servers which LHCb uses to buffer data while performing real-time alignment and calibration, and so on. Even if such a processing could be economically affordable, the relevant power consumption would be hard to defend. Additionally, LHCb's new data processing facilities (HLT1, HLT2, Event Building and data Buffering) are limited to 2.0~MW, which would have been insufficient. Even with the gain in compute power of more modern processors we would have fallen short here.

Looking further ahead, LHCb is currently proposing to build a second upgrade~\cite{LHCb-PII-EoI} in the 2030s which would roughly speaking increase the data rate and hence processing complexity by another order of magnitude. It is then clear that without
significant gains in energy efficiency, the situation will quickly become untenable.

Fortunately, as we have documented, such significant gains have been achieved over the past years through a combination
of software optimization, physics improvements, and exploitation of new processing architectures. The overall gain
of one order of magnitude in energy efficiency is split fairly evenly between these different components, although the
rethinking of our algorithms and software framework to support SIMD and SIMT processing paradigms is undoubtedly the single
most important aspect contributing to this improvement. The financial savings associated with this improvement is a modest 76~kEUR per year.\footnote{If we assume $6\cdot 10^6$ seconds of datataking per year and the cost of $0.042$ Euro per kWh for electricity which CERN pays.~\cite{cernelectricitycost}} %, we note that this saving is in the same ballpark as the price of 200 GeForce~RTX~2080~Ti GPU cards required to implement HLT1. It's also roughly similar to the annual maintenance and overhaul budget for LHCb's real-time computing hardware. 
The situation with HLT2, where much larger potential power savings are available, is complicated
by the necessity to reuse LHCb`s existing Run~2 servers due to financial constraints. On one hand, optimizing for these legacy machines would give better efficiency, on the other hand it is not clear how much of these efforts can be ported over to more modern architectures. Some of the machines here are reaching 10 years of service. We expect analogous improvements and savings
there on the timescale of the second LHCb upgrade, when these machines can be replaced by more modern architectures.

As encouraging as these results are, a word of caution is required with regards to the future. The improvements to our
software in order to exploit SIMD and SIMT processing paradigms are not something we can expect to easily repeat for the
second upgrade. While further incremental improvements are certainly to be expected, gaining factors from purely
software engineering improvements is unlikely. Another way to look at this is that the optimized HLT1 codebase saturates 
the FLOP capacity of current CPU architectures. The situation is not quite the same for GPUs, where the available theoretical
FLOPs remain far larger than those used by the current GPU HLT1 implementation. However the fragmented nature of GPU
processors and their cores which specialize in one kind of calculation or another mean that fully exploiting the theoretical
computing capacity for complex HEP workflows will be difficult if not impossible. Gaining another order of magnitude in
energy efficiency in time for the second LHCb upgrade will therefore require a great deal of patient optimization of each
component in our processing chain. 

\section*{Acknowledgements}
The authors would like to thank the LHCb computing and simulation teams for their support and for producing the simulated LHCb samples used to benchmark the performance of RTA software. VVG and AH acknowledge support of the European Research Council under Consolidator grant RECEPT 724777. CF is grateful for the support of UKRI (United Kingdom).

%
% BibTeX or Biber users please use (the style is already called in the class, ensure that the "woc.bst" style is in your local directory)
\bibliography{main}
%\begin{thebibliography}{}
%\bibitem{LHCb-FIGURE-2020-016}
%LHCb Collaboration, Figure document 2020-16, (2020)
%Book Author, \textit{Book title} (Publisher, place, year) page numbers
% etc
%\end{thebibliography}

\end{document}